# A DFT+$U$ study on the contribution of 4f electrons to oxygen vacancy formation and migration in Ln-doped $CeO_2$

Musa Alaydrus[1], Mamoru Sakaue[1,2], and Hideaki Kasai[1,3,4*]

[1]Department of Applied Physics, Graduate School of Engineering, Osaka University,
2-1 Yamadaoka Suita, Osaka, 565-0871, Japan
[2]Center for International Affairs, Graduate School of Engineering, Osaka University,
2-1 Yamadaoka, Suita, Osaka 565-0871, Japan
[3]National Institute of Technology, Akashi College, 679-3 Nishioka,
Uozumi-cho, Akashi, Hyogo 674-8501, Japan
[4]Institute of Industrial Science, The University of Tokyo, 4-6-1 Komaba, Meguro-ku, Tokyo,
153-8505, Japan
[*]E-mail: kasai@dyn.ap.eng.osaka-u.ac.jp

Keywords: density functional theory, solid electrolytes, oxygen ion conduction, 4f electrons

**Abstract:**
Rare-earth doped form, ceria ($CeO_2$) is of interest as a potential candidate for solid oxide fuel cells (SOFCs) because of its relatively high oxygen ion conductivity at temperatures below 600 °C. At the present time, computational chemistry has reached a certain maturity which allows prediction of materials properties that are difficult to observe experimentally. However, understanding of the roles of dopants on the oxygen ion conduction in $CeO_2$ is still incomplete for quantitatively reliable analysis due to strong electron correlation of 4f electrons. In this study, density functional theory calculations with Hubbard $U$ corrections are conducted to discuss ionic/covalent interactions in rare-earth-doped $CeO_2$ and their consequences to oxygen ion conduction. The study suggests that the variable occupancy of empty 4f orbitals is important typically for early Ln elements to produce the covalent interactions that essentially affect formation and migration of oxygen vacancies. This finding is important in understanding the factors responsible for oxygen ion diffusion in doped $CeO_2$.

**Introduction**

Ceria ($CeO_2$)-based materials have been suggested as promising solid electrolyte (SE) material candidates for realizing solid oxide fuel cells (SOFCs) with intermediate to low operating temperature, viz., below 600°C.[1] The high oxygen ion conduction in $CeO_2$-based SEs is related to the cubic fluorite type crystal which has a fairly open space within its octahedral holes.[2] Aliovalent doping, in particular rare-earth (RE) elements, promotes a large concentration of charge compensating oxygen vacancies, resulting in high mobile oxygen concentrations for oxygen diffusion through vacancy diffusion mechanism.[3-6]

Ionic conductivity of oxide conductors is governed by Arrhenius equation where the conductivity $\sigma$ is proportional to the exponential of the activation energy $E_a$, the energy required for moving the oxide ions, $\sigma \propto \exp(-E_a/k_B T)$. Here, $T$ and $k_B$ denote the operating temperature and Boltzmann's constant, respectively. This relation shows that it is crucial to minimize $E_a$ to maximize the conductivity in low temperatures.[5, 6] Further, formation of defect associates of oxygen vacancies with dopants restricts the number of mobile oxygen and hence contributes to a high $E_a$.

Table 1. Experimental bulk activation energies, $E_a$, of doped $CeO_2$. [a]ref. [7] [b]ref. [3]; [c]ref. [8]; [d]ref. [9]; [e]ref. [1].

| Dopants | Ionic Radius[a] (Å) | $E_a$ (eV) |
|---|---|---|
| La | 1.16 | 0.73[b] |
| Pr | 1.13 | 0.70[c] |
| Nd | 1.11 | 0.69[b], 0.72[d] |
| Sm | 1.07 | 0.66[e], 0.69[d] |
| Gd | 1.05 | 0.64[e], 0.70[b,d] |
| Y | 1.02 | 0.74[d], 0.78[b] |
| Yb | 0.99 | 0.85[b] |

There were many studies related to $CeO_2$ based SEs, both experiments and theories[1, 3, 9-30] including effects of co-doping as well as lattice strain effects on oxygen ion conduction in doped $CeO_2$.[9, 12, 17, 20-23, 26] Throughout the course of understanding the mechanism of oxygen ion conduction in RE-doped $CeO_2$, the focus is often to address that the performance of SEs, through the activation energy of vacancy diffusion mechanism. It is generally considered about $CeO_2$-based SEs that small ionic radii mismatch between the host and dopant cations results in lower activation energies.[5, 6] However, the experimentally found relation between activation energies and dopant ionic radii shows discrepancy from this simple theory as shown in **Table 1**. While the ionic radii of $Ce^{4+}$ and $Ce^{3+}$ are 0.97 Å and 1.14 Å,[7] respectively, the smaller mismatch with $Ce^{3+}$ rather than $Ce^{4+}$, which should be the essential oxidation number as the host, gives lower activation energies except for La. The trend among Nd, Sm, and Gd is unclear and within the variance caused by the difference in sample preparation. These facts indicate existence of unknown complicated mechanisms determining the activation energies.

Density functional theory (DFT)[31, 32] is a powerful tool for analyzing and predicting properties of various materials, however, it has essential difficulties in treatment of lanthanide (Ln) and heavier elements because of the strong correlations between 4f electrons. So far, the existing studies of doped $CeO_2$ based on DFT calculations have avoided essential discussions on the particular roles of 4f electrons of dopants by assuming that the 4f electrons are too localized to participate in the bonding.[12, 15, 20, 23] These facts have prevented detailed understanding, especially, in elucidating the effects of Ln doping in Ln-doped $CeO_2$. In this paper we report our systematic study on the atomic and electronic properties in M-doped $CeO_2$ (i.e. M = La, Pr, Nd, Pm, Sm, Eu, and Gd, as part of Ln elements, as well as Y chosen from non-Ln rare-earth elements as a reference) focusing on the contribution of 4f-states to cation-anion interactions. Note that Pm is a radioactive element and the reason of its inclusion in this study is purely academic. The study was conducted by means of DFT calculations and focusing on the doping effects on the stable configurations and the migration energies of oxygen ion in doped $CeO_2$. To consider the localization effects introduced by 4f electrons of all Ln elements, the strong electron correlations were treated within the generalized gradient approximation (GGA) with the Hubbard $U$ correction.[33] Electronic and bonding properties are discussed to semi-quantitatively explain the 4f-states contributions to the cation-anion interactions, in particular, in its relation with oxygen ion migration in doped $CeO_2$.

**Computational details**

The calculations were performed with the Vienna *Ab initio* Simulation Package (VASP)[34, 35] and visualized using Visualization for Electronic and STructural Analysis (VESTA).[36] The interactions between ions and electrons were described by the projector augmented wave (PAW) method[37] and the exchange correlation effects were described by the generalized gradient approximation (GGA) using the Perdew-Burke-Ernzerhof (PBE) functional for solids.[38] The oxygen atoms have been described by $2s^2 2p^4$ valence electrons and $5s^2 5p^6 6s^2 5d^1 4f^{n-1}$ for Ln elements ($n$ = number of occupied 4f orbitals). Yttrium was described by $4s^2 4p^6 5s^2 4d^1$ valence electrons. For reference, we also performed calculations where the Ln 4f electrons were considered as core electrons (4fcore), $5s^2 5p^6 6s^2 5d^1$ for Ce (only for two reduced Ce ions as virtual dopants in the presence of an oxygen vacancy), Pr, Nd, Pm, and Sm atoms and $5p^6 6s^2 5d^1$ for Gd. The strong correlation effects introduced by 4f electrons of all Ln elements were treated within the GGA with the Hubbard $U$ correction formulated by Duradev et al.[39] For Ln dopants, the optimum on-site Coulomb interaction, $U_{eff@Ln}$, was fitted by scanning from 0 to 6 eV at interval of 0.5 eV so as to agree with either experimental 4f energy band positions relative to $O_{2p}$-valence band ($O_{2p}$-VB) and $Ce_{5d}$-conductance band ($Ce_{5d}$-CB) as in Pr-doped $CeO_2$[16] and/or with Hybrid calculations[40] for the rest (due to lack of available experimental spectroscopy data). The fitted $U_{eff@Ln}$ values are as follows: $U_{eff@La}$ = 4.5 eV, $U_{eff@Pr}$ = 4 eV, $U_{eff@Nd}$ = 3.5 eV, $U_{eff@Pm}$ = 3.5 eV, $U_{eff@Sm}$ = 4 eV, $U_{eff@Eu}$ = 4 eV, $U_{eff@Gd}$ = 4 eV, and $U_{eff@Yb}$ = 5.5 eV. Here the results of the hybrid functional calculations are provided in the supporting information. The estimation of single $U_{eff@Ln}$ also has been made possible due to a fact that the profile of 4f energy band positions relative to $m$p-VB and 5d-CB follow a universal behavior (zigzag pattern) in various compounds as summarized in the work of Rogers et al.[41] In all calculations, $U_{eff@Ce}$ is set to 5 eV.

The supercell was constructed by repetition of the unit cell of ceria with space group of $Fm\bar{3}m$ and consists of 96 atoms for pristine $CeO_2$ (see Figure 1). A plane wave basis set up to an energy cutoff of 520 eV as well as a 2×2×2 *k*-points of Monkhorst-Pack grids[42] have been used for geometry optimization while for the total energy as well as electronic density of states (DOS) calculations we applied 4×4×4 *k*-points of Monkhorst-Pack grids. Full optimization of the atomic positions and lattice parameters were performed by minimizing the Hellman-Feynman forces[43] acting on each ion to be less than 0.02 eV/Å. Cation-anion bonding analysis based on crystal orbital Hamilton population (COHP) was done by utilizing Local-Orbital Basis Suite Towards Electronic Structure (LOBSTER-2.0.0) program.[44-46] Finally, we address the activation energy $E_a$ as migration energy $\Delta E_m$ and were calculated by climbing image nudge elastic band (CI-NEB) method.[47] The selection of migration paths considered in this study were chosen based on the position of the dopants within the first coordination shell of the migration paths, as has been reported in our previous study,[23] as illustrated in **Figure 1**.

**Results and discussion**
**Pristine and reduced $CeO_2$**

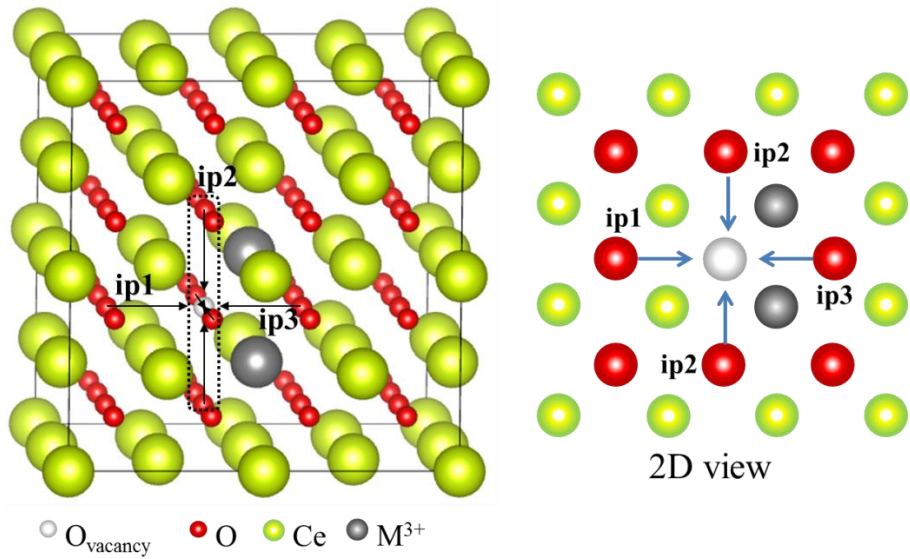

**Figure 1.** (Color online). Atomic configuration and oxygen migration pathways of oxygen-deficient and M-doped ceria. (a) Supercell of reduced ceria ($CeO_{1.97}$ and $M_{0.06}Ce_{0.94}O_{1.97}$ with M = La, Pr, Nd, Pm, Sm, Eu. Gd, and Y). Arrows ip1, ip2, and ip3 indicate probable individual migration paths, along which the dopant-vacancy coordination within the first coordination shells. (b) The simplified 2-dimensional view of the corresponding migration paths showed in (a). $M^{3+}$ indicates a reduced cation, $Ce^{3+}$ or $M^{3+}$. In ip2, there are four equivalent migration paths as indicated by dotted rectangle in the supercell.

$CeO_2$ adopts the cubic fluorite ($CaF_2$) structure type (see Figure 1a). Each Ce atom surrounded by eight O atoms while each O atom has four coordination numbers with Ce atoms. The calculated lattice constant *a* of pristine $CeO_2$ is 5.429 Å in a good accordance with the experimental value of 5.411 Å.[14] The nearest neighbor Ce-O bond is 2.35 Å an exact sum of the two ionic radii of $Ce^{4+}$ (0.97 Å) and $O^{2-}$ (1.38 Å) according to Shannon effective ionic radii [7].

The total density of states (DOS) and projected DOS plots along with Ce-O COHP curve are shown in **Figure 2**. The projected DOS shows that the states above the Fermi energy (at 2 – 4 eV) are primarily showing 4f character. The states in -4 – 0 eV are primarily composed of $O_{2p}$ states with a considerable admixture of the $Ce_{4f}$ orbitals at the upper band and $Ce_{5d}$ orbitals at the lower band. The COHP curve reveals that the mixing states in -4 – 0 eV are Ce-O bonding due to orbital hybridization (positive value means bonding while negative value is anti-bonding). This indicates existence of covalent interactions in this particular ionic solid system. The integrated COHP of single Ce-O pair is -3.48 eV. The obtained COHP results in pristine $CeO_2$, i.e. the shape showing bonding and antibonding interactions that are essential for chemical bonding analysis, correctly reproduce previously reported COHP analysis of $CeO_2$ with Linear Muffin-Tin Orbital (LMTO).[48] The absolute value of integrated COHP gives indication of the covalent bond strength. Though, in pristine $CeO_2$, on its own, is not particularly informative, but the changes in it under doping turn out to be quite indicative.

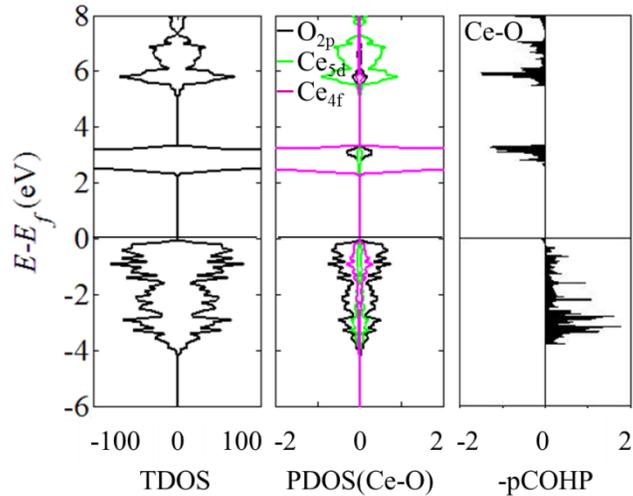

**Figure 2.** (Color online). Total density of states (DOS), projected DOS on neighboring Ce and O ions, and the corresponding projected COHP curves of single Ce-O pair.

In the cubic phase of pure $CeO_2$ with oxygen vacancy, there are three possible direct oxygen ion migration pathways as shown in the left figure of Figure 1. The presence of defects, i.e. oxygen vacancy and/or doping, promotes lattice deformation in the fluorite supercell as has been discussed by Marrochelli *et al.*[17, 18] The origin of lattice deformation is a result of two competing mechanisms: (i) change in ionic radii of reduced or substituted Ce ions, i.e. $Ce^{4+}$ to $Ce^{3+}$ or $Ce^{4+}$ substituted by $M^{3+}$, respectively; and (ii) decrease in lattice volume with the formation of charge compensating oxygen vacancies due to doping. The lattice *a* expanded to 5.441 Å in the reduced cell of $CeO_{2-x}$. Two electrons per vacancy are redistributed and localize onto and hence reduce the two nearest Ce ions, as shown in **Figure 3**. These localized charges modify the radius of the two reduced Ce ions causing the lattice expansion.

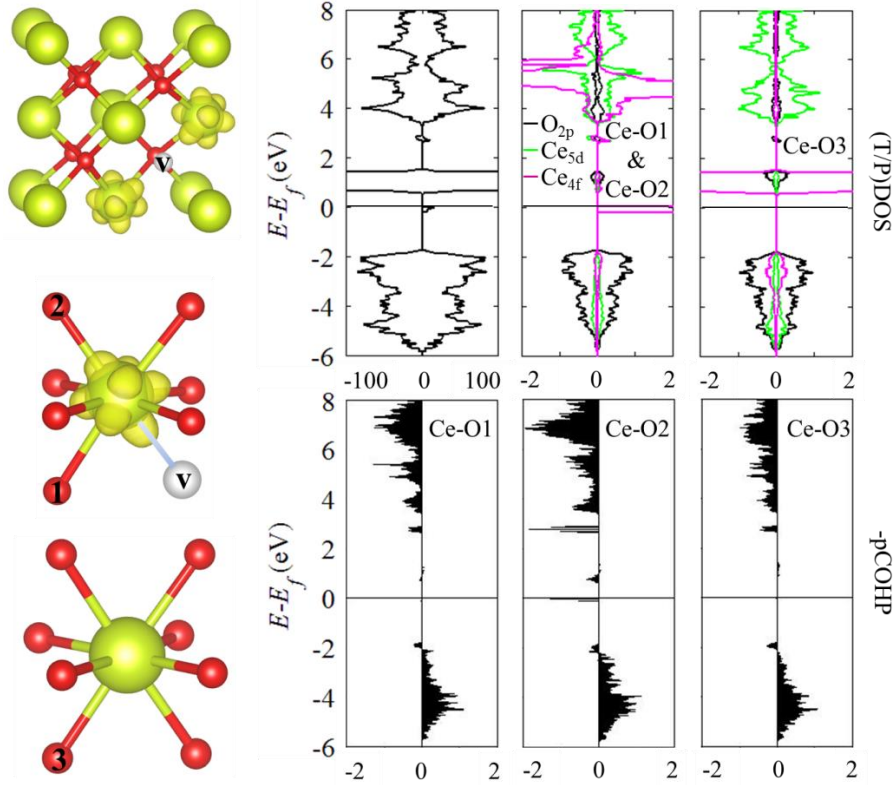

**Figure 3.** Spin-density of CeO$_{2-x}$ and the corresponding total DOS, projected DOS, and COHP curves of Ce-O bonds. The isosurface value of the spin-density is 0.058 e Å$^{-3}$.

As it was in pristine CeO$_2$, the admixture of O$_{2p}$ states and Ce$_{4f}$ and Ce$_{5d}$ states in CeO$_{2-x}$ below the Fermi energy is observed in Figure 3 suggests covalent interactions. The projected DOS and COHP curves for three different Ce-O bonds are labeled by number 1, 2, and 3 as shown in the figure. The localized charges which are indicated by the emerging defect state just below $E_f$ show anti-bonding character (see Ce-O2 COHP curve in Figure 4). In Ce-O1 and Ce-O2 bonds, Ce ion is in reduced state, 3+, while in Ce-O3 bond it is in tetravalent state. The integrated Ce-O COHP pairs are -3.07 eV (Ce-O1, 2.40 Å), -3.65 eV (Ce-O2, 2.31 Å), and -3.69 eV (Ce-O3, 2.32 Å). As indicative from the integrated COHP, the interionic distance between Ce and O ions as well as the number of empty orbitals before interactions, i.e. Ce$^{4+}$ is 7 and Ce$^{3+}$ is 6, determine the magnitude of the interactions. More quantitative discussion by comparison with the doped systems will be conducted in the next section. In both cases, pristine and reduced CeO$_2$, the partially occupied Ce$_{4f}$ and Ce$_{5d}$ orbitals are strongly involved in Ce-O bonds. The calculated migration energy $\Delta E_m$ is 0.442 eV and the detail will be discussed by comparison with the doped systems in the next section.

**Aliovalent doped CeO$_2$**

In the presence of trivalent dopants, the lattice of CeO$_2$ systems undergoes deformation almost linearly dependent on the radius of the dopant ions (see **Figure 4**). Small radii result in lattice contraction while larger radii promote lattice expansion. The closest lattice constant Gd doping to pristine $CeO_2$ indicates the minimum lattice distortion among the dopants. This is due to the expansion originated from Gd ions is cancelled out by the decrease in the volume due to vacancy formation. Still in Figure 4, in the first coordination shell of dopant-vacancy configuration, that the bond-lengths between Ce and $O_{mig}$ as well as M and $O_{mig}$ also show dependency on the dopant ionic radius. Here, $O_{mig}$ is the migrating O ion. Further, we also investigated the effect of variable occupancy of Ln 4f electrons on the lattice deformation by comparison between the 4fval and 4fcore models. For early Ln elements, the lattice experiences contraction with $U_{eff@Ln}$ = 0 eV. As $U_{eff@Ln}$ increases, the lattice constants expand. The change of the lattice with the applied $U_{eff@Ln}$ can be directly related to the electron localization on the Ln ions. However, when the 4f orbitals almost and/or half-fully occupied, the expansion with the increase of $U_{eff@Ln}$ is suppressed or cancelled out. Both 4fval with fitted $U_{eff@Ln}$ and 4fcore models result in matched lattice constant as shown in Figure 4b. For example, in Pr doping, the 4fcore calculation leads to $a$ = 5.439 Å which is matched with the 4fval calculation with the fitted $U_{eff@Pr}$ = 4 eV. This applies for reduced and all other doped $CeO_2$.

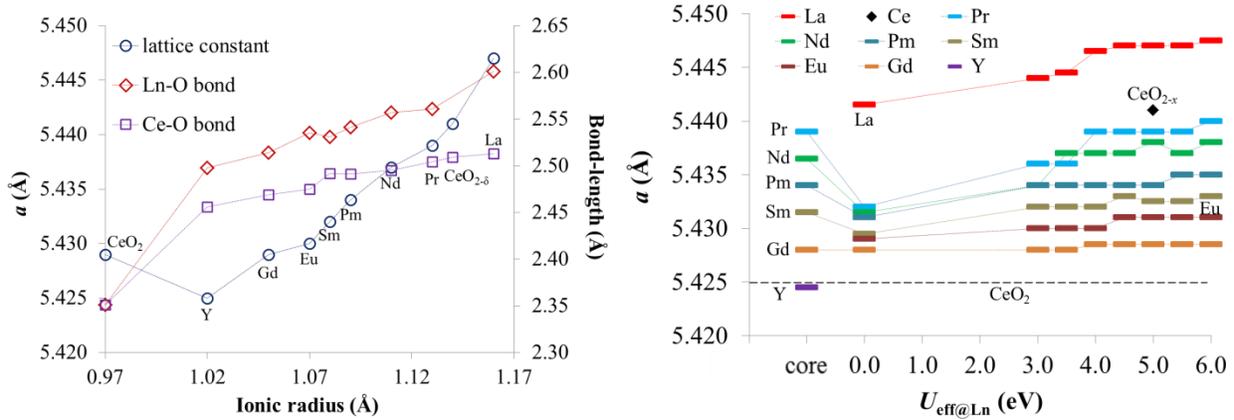

**Figure 4.** (a) The calculated lattice constants of doped $CeO_2$ (left ordinate) and cation-$O_{mig}$ bond-lengths (right ordinate) of doped $CeO_2$ as functions of dopant ionic radii (according to Shannon effective ionic radii [7]) for Ln-doped systems with fitted $U_{eff@Ln}$, Y-doped systems, and the pristine system. The results for the 4f-core model show no significant difference from the 4fval model with the fitted $U_{eff@Ln}$ (not shown). (b) The calculated lattice constants of doped $CeO_2$ as functions of $U_{eff@Ln}$, where "core" indicates values by the 4fcore model. The horizontal dashed line indicates the value for pristine $CeO_2$, where $U_{eff@Ce}$ is set to 5 eV. Note that Y has no 4f orbitals on its outer shell.

The relative energies between the initial and final states of migration path ip1, $\Delta E_{ist,fst(1)} = \Delta E_{fst(1)} - \Delta E_{ist}$, for various types of M doping systems are shown in **Figure 5**. $\Delta E_{ist}$ and $\Delta E_{fst}$ denote total energies at initial and final states that are (meta-)stable configurations, respectively. Positive value means the oxygen vacancy prefers to coordinate with the dopants. The more positive the values the more the oxygen vacancy tends to associate with the dopants. The 4fval calculations suggest that only when doped with La the oxygen vacancy favors a site next to Ce ions with $\Delta E_{ist,fst(1)}$ = -0.16 eV. Meanwhile, the 4fcore approach predicts linear relationship between $\Delta E_{ist,fst(1)}$ and dopants ionic radii as well as that oxygen vacancy tends to form at the second nearest neighbor to the trivalent ions in Pr doping and $CeO_{2-x}$ (i.e. the reduced Ce ions,

$Ce^{3+}$, was described by having the 4f electrons at the core) cases. This preference in oxygen vacancy formation obtained by the 4fval calculations agrees with an EXAFS analysis[19] that suggested the preference around Pr and Ce ions. Even though the difference in the relative energies, $\Delta E_{ist,fst(1)}$, between the two treatments is small, this subtle change of the preferential site for oxygen vacancy formation is an evidence of the importance of variable occupancy of dopant 4f orbitals.

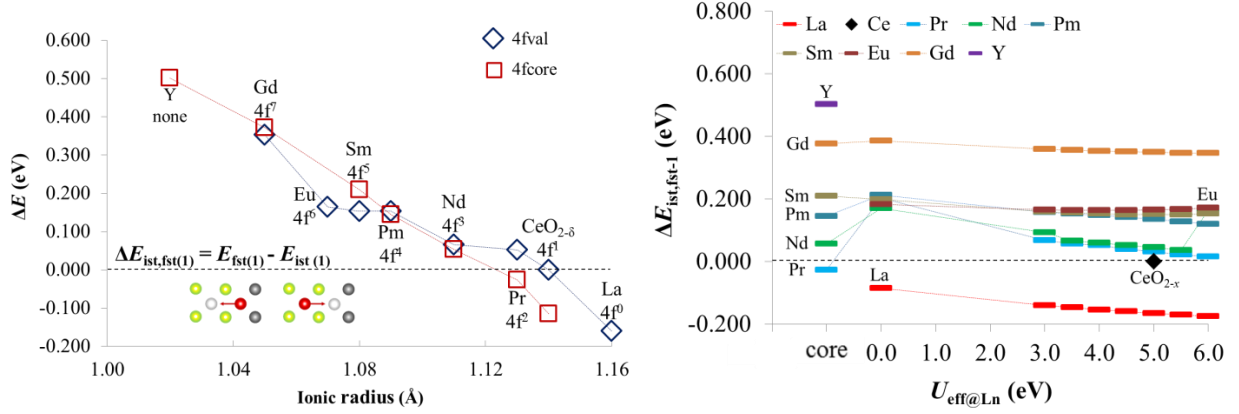

**Figure 5.** Energy difference between the initial and final states of ip1, $\Delta E_{ist,fst(1)}$. (a) $\Delta E_{ist,fst(1)}$ as functions of the dopant ionic radius, where the element names and numbers of occupied 4f orbitals are attached to the corresponding data points. (b) $\Delta E_{ist,fst(1)}$ as a function of $U_{eff@Ln}$, where "core" indicates values by the 4fcore model.

Formation energy of oxygen vacancy, $E_f^{vac}$, is defined as the energy needed to remove an oxygen atom per supercell in the form of half oxygen molecule from $CeO_2$ in the neutral condition and formulates as

$$E_f^{vac} = E(CeO_{1.97}) + \tfrac{1}{2}E(O_2) - E(CeO_2) \qquad (1)$$

for undoped $CeO_2$ ($CeO_{2-\delta}$), and

$$E_f^{vac} = E(M_{0.06}Ce_{0.94}O_{1.97}) + \tfrac{1}{2}E(O_2) - E(M_{0.06}Ce_{0.94}O_2) \qquad (2)$$

for doped $CeO_2$. Here $E(O_2)$ is the total energy of an oxygen molecule. Low $E_f^{vac}$ translates to high possibility of oxygen vacancy formation. The calculated $E_f^{vac}$ with the 4fval and 4fcore models for various doped $CeO_2$ are shown in **Figure 6**. In the 4fcore model $E_f^{vac}$ shows direct dependency on the dopant ionic radius. Meanwhile, in the 4fval model, only doping with Pm or heavier elements show linear dependency on the dopant ionic radius. La doping falls in between Pm and Sm doping, while Pr and Nd doping have higher $E_f^{vac}$. Note that the calculated $E_f^{vac}$ for Pr doping with the fitted $U_{eff@Pr} = 4$ eV, 1.64 eV, while in experiment reported by Kim *et al.*[8] was 1.90 eV. Although the calculated $E_f^{vac}$ with 4fval model underestimates the experiment value, it gives better agreement than with 4fcore model which results in negative $E_f^{vac}$.

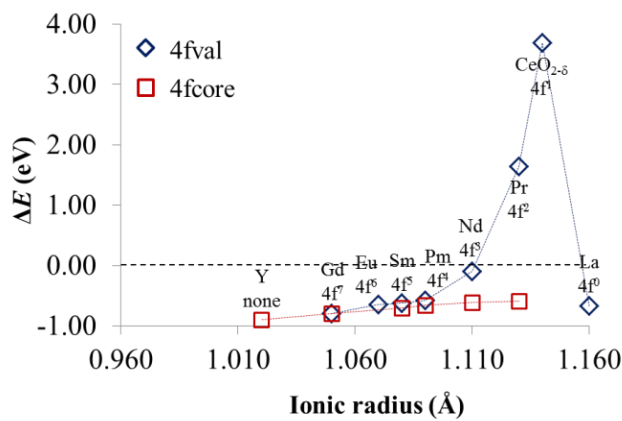

**Figure 6.** The calculated oxygen vacancy formation energy by the 4fval model with the fitted $U_{\text{eff@Ln}}$ (diamond) and the 4fcore model (square) as a function of dopant ionic radius.

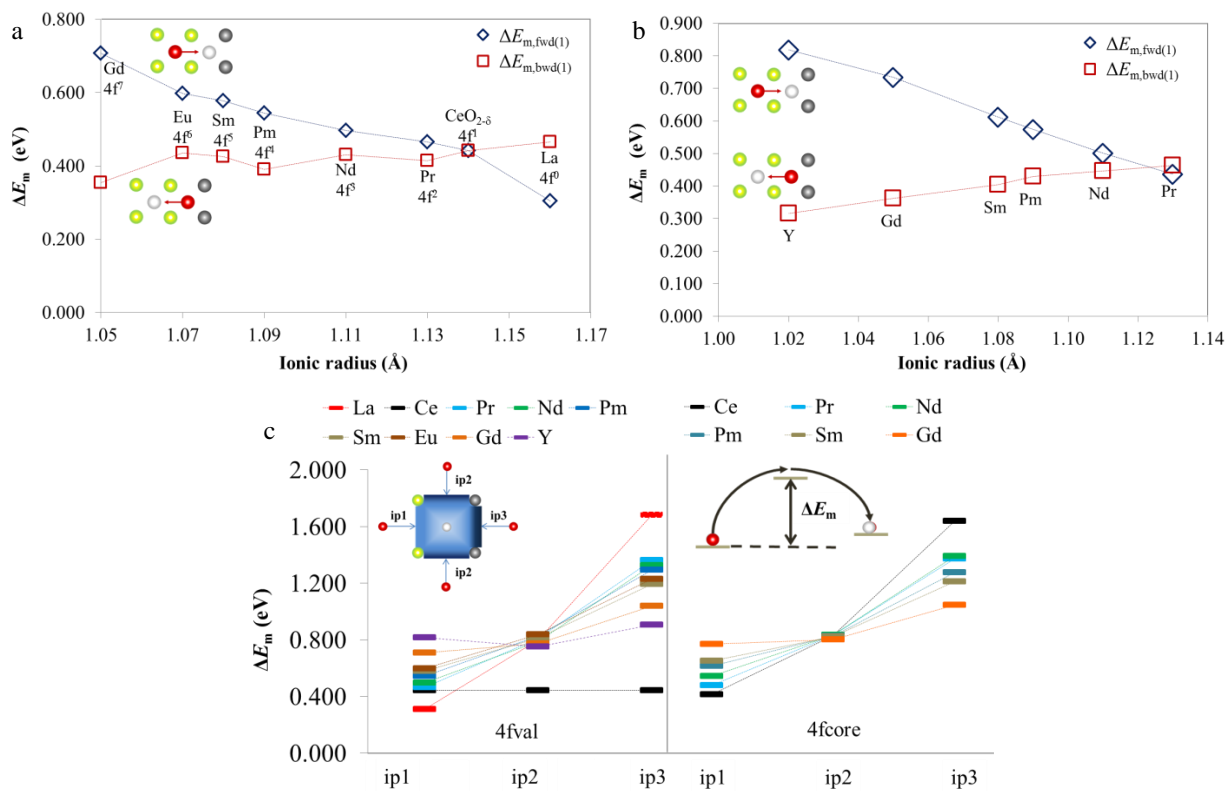

**Figure 7.** Forward and backward oxygen migration energies in ip1. (a) 4fval model and (b) 4core model with insets showing the corresponding migration paths in the first coordination shell. (c) Forward migration energies of ip1, ip2, and ip3 for all doped $CeO_2$ by the 4fval model (left pane) and the 4fcore model (right pane). The inset of the left pane shows a 2-dimensional view of oxygen migrations. The inset of the right pane shows the definition of barrier height. "Ce" indicates results for reduced $CeO_2$.

The connection between $\Delta E_{\text{ist,fst(1)}}$ and the migration energies $\Delta E_{\text{m}}$ of each doped systems is immediately seen in Figure 5 and **Figure 7** where it is found that higher $\Delta E_{\text{ist,fst(1)}}$ gives higher $\Delta E_{\text{m}}$. Note that we evaluate the migration energies by differentiating forward and backward (in the reverse direction) migrations. In forward migration, the migrating oxygen, $O_{mig}$, is to break the bond with Ce ions at the side while in backward migration, $O_{mig}$ to break the bond with M dopants at the side. For example, in Pr-doped $CeO_2$, $\Delta E_{ist,fst(1)}$ is 0.039 eV and $\Delta E_m$ are 0.466 and 0.414 eV for forward and backward migrations, respectively, while in Gd-doped $CeO_2$ with $\Delta E_{ist,fst(1)} = 0.353$ eV has $\Delta E_m = 0.708$ and 0.355 eV. In these particular systems, where $\Delta E_{ist,fst(1)}$ are positive, the oxygen vacancy prefers to coordinate with the dopants, and therefore, the migrating oxygen has higher probability to first meet the forward migration barrier which gets higher as the energy difference gets larger. These results indicate that in order to minimize $\Delta E_m$ we need dopants that can balance the two stable configurations.

Figure 7c summarizes the forward migration energies of all the individual paths, ip1, ip2, and ip3 for 4fval (left panel) and 4fcore (right panel) models. Both models result in similar trends in predicting the migration energies in all three directions, i.e. ip1, ip2, and ip3, except for reduced $CeO_2$ by the 4fcore model. This unphysical difference among the individual paths in reduced $CeO_2$ is caused by the broken isotropy due to the fixed 4f electron occupancy in the "dopant" site, i.e. reduced Ce ions. The important point can be deduced from these results is that, even though early Ln elements produce a larger variation of $\Delta E_m$ among the three individual paths, so that ip1 becomes more favorable than the others, dopants with smaller radii still tend to trap the oxygen vacancy strongly so that the vacancy hardly be released even via the most favorable path. This is in-line with the previous discussion on $\Delta E_{ist,fst(1)}$ that $O_{mig}$ has higher probability to first overcome the forward migration barrier which gets higher as the energy difference gets larger.

Up to here, this study suggests that: (i) geometrical structures of (meta-)stable configurations are predominately determined by host-dopant ionic radii matching; and (ii) energetics of formation and migration is still governed by host-dopant ionic radii matching but considerably modified by variable occupancy of 4f electrons. For instance, $E_f^{vac}$ of La doping lies in between Eu and Gd doping. The 4fval model calculations show that $\Delta E_{ist,fst(1)}$ for Pr and Nd doping and that for Pm, Sm, and Eu doping form groups of similar values. However, the 4fcore model results in a linear dependency on the dopant ionic radius. Thus there might be additional factor(s) related to 4f electrons. In order to elucidate this, in the following discussion, further elaboration on the contributions of 4f-states in the cation-anion interactions based on the electronic properties and bonding analysis following the same procedure as for pristine and reduced $CeO_2$ is presented.

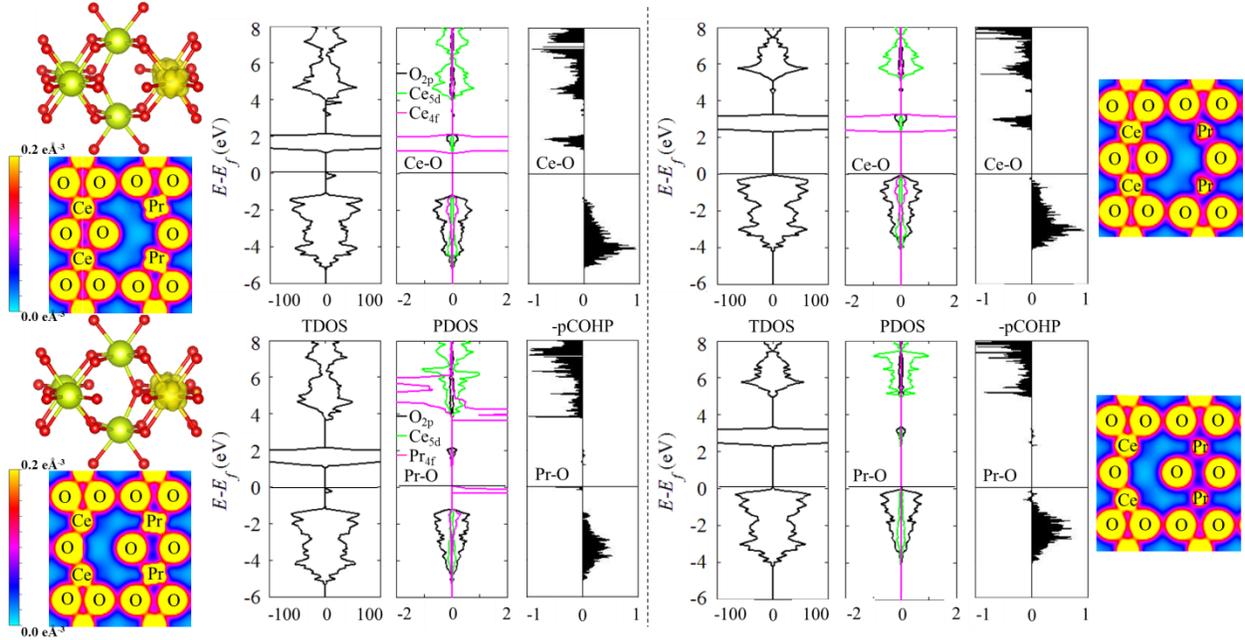

**Figure 8.** Spin density, charge density, total and projected DOS of Pr-doped $CeO_2$, as well as the corresponding cation-$O_{mig}$ projected COHP of ip1 for 4fval model (left panel) and 4fcore model (right panel) separated by vertical dashed line. Properties at the initial and final states of ip1 are positioned at upper and lower sides, respectively. The charge density plots are obtained within energy slice containing only the levels between $E_f$ and 5 eV and 4 eV below $E_f$ for the 4fval and 4fcore models, respectively. The color bar range is from 0 to 0.2 $eÅ^{-3}$.

In **Figure 8**, as an illustration, the electronic properties in terms of spin density, charge density on the $O_{2p}$ band energy range, and the corresponding Ce-$O_{mig}$ and Pr-$O_{mig}$ projected COHP curves at initial and final states of ip1, respectively, are presented. In the 4fval model we observe considerable admixture between $O_{2p}$ band and $Ce_{4f}$ and $Ce_{5d}$ bands at initial state and between $O_{2p}$ band and $Pr_{4f}$ and $Pr_{5d}$ bands at final state of ip1. From the spin density it is seen that the excess charge due to removal of an oxygen atom in the charge neutral condition are essentially localized on the Pr ions as indicated as well in the PDOS. pCOHP curve shows that this defect state has anti-bonding character, similar to the case of $CeO_{2-x}$. On the other hand, in the 4fcore model, only $Pr_{5d}$ band presents while the empty 4f states of Pr ions are completely removed from the electronic description. As it is seen in the charge density plot at the top right corner, the electron distribution of Pr ions is isotropic reflecting the characteristic of ionic bonding. At the initial state, both 4fval and 4fcore models show similar bonding properties between Ce-$O_{mig}$ except for the shift of Fermi energy due to the defect state on Pr. The discussion for other doping systems can be built following the same development as in reduced and Pr-doped $CeO_2$; and the corresponding results can be found in the supporting information.

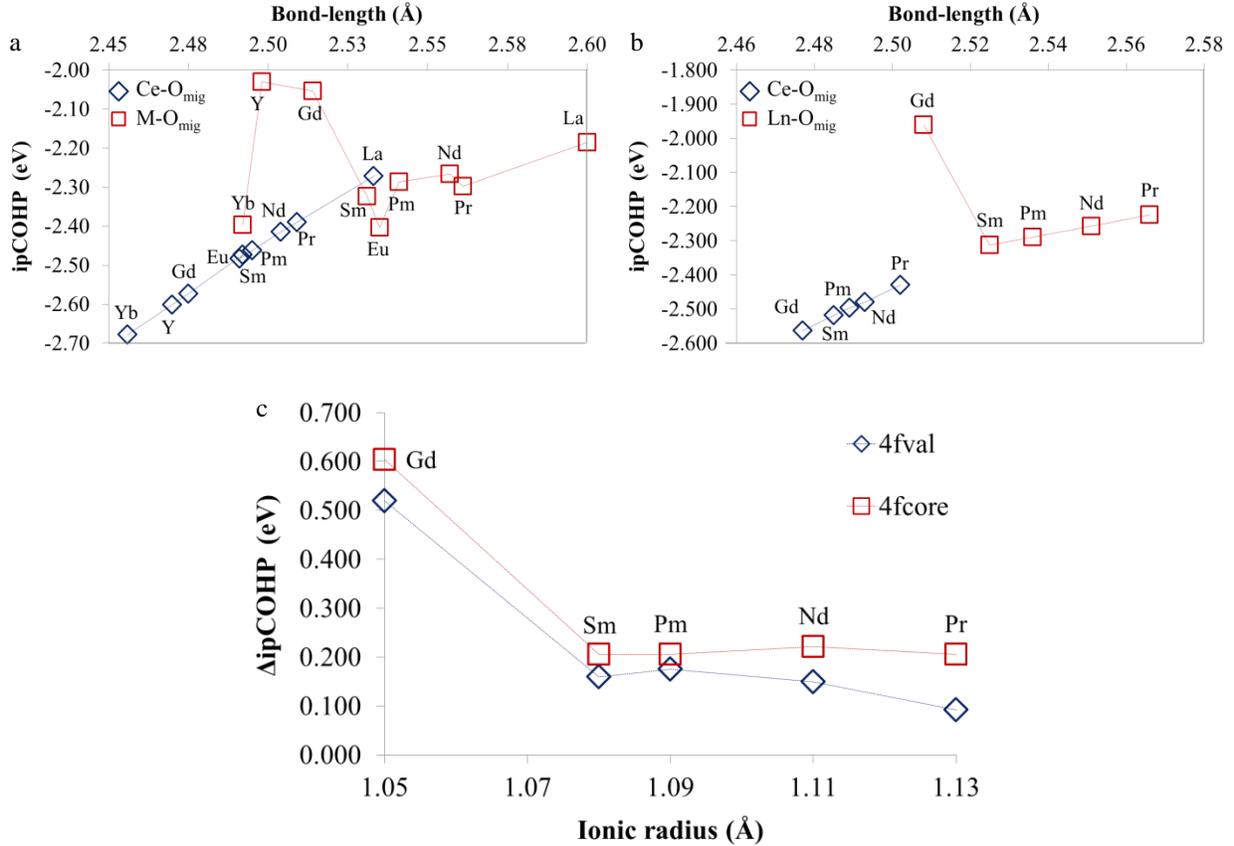

**Figure 9.** Integrated COHP of Ce-O$_{mig}$ and M-O$_{mig}$ bonds at the initial and final states of the forward and backward migration of ip1, for M-doped CeO$_2$: (a) 4fval model with fitted $U_{eff@Ln}$, (b) 4fcore model, and (c) the difference in the integrated pCOHP, $\Delta$ipCOHP = pCOHP(M-O$_{mig}$) − pCOHP(Ce-O$_{mig}$).

**Figure 9** shows the relation between cation-O$_{mig}$ bond-length and the integrated pCOHP as well as the difference in integrated pCOHP, $\Delta$ipCOHP = pCOHP(Ln-O$_{mig}$) − pCOHP(Ce-O$_{mig}$), at the initial state of the forward (Ce-O$_{mig}$) and backward (M-O$_{mig}$) migration of ip1. It is seen that in the forward migration there is linear relationship between Ce-O$_{mig}$ bond-length and the magnitude of ipCOHP for both 4fval and 4fcore models. However, in the backward migration, it is clearly seen that in the linearity is highly modulated in the 4fval model while it is preserved in the 4fcore model.

One way to rationalize all the above observed phenomena is to revisit the fundamental properties of pristine CeO$_2$. In the ideal theory of oxidation states of ionic systems, CeO$_2$ is composed of Ce$^{4+}$ and O$^{2-}$ ions. It means that each Ce atom donates four of its electrons (i.e. $6s^2 5d^1 4f^1$) to two O atoms (the p orbital of O ions become fully occupied). Starting from this point of view one may initially guess that CeO$_2$ should have empty 4f and 5d orbitals and therefore it should not be observed in the projected DOS. However, DFT+$U$ calculations reveal the admixture of O$_{2p}$ band with Ce$_{4f}$ and Ce$_{5d}$ bands (see Figure 3) which results in fractional atomic charges of ions as understood from the charge population analysis like Bader charge analysis,[49, 50] i.e. Ce is 2.27 and O -1.14 electrons per ion. This is caused by the overlap of wave functions of $Ce_{4f}$ and $Ce_{5d}$ orbitals[51] resulting in hybridization between $O_{2p}$ and $Ce_{4f}$ orbitals to form delocalized f bands in addition to strongly localized f bands. Therefore, the originally empty Ce orbitals, i.e. $Ce_{4f}$ and $Ce_{5d}$, play an important role in the covalent interactions as confirmed by the COHP analysis. The same argument applies to the Ln doped $CeO_2$. One of important characteristics is change in the bonding due to doping which causes lattice deformation. As shown in Figure 9, in various doped $CeO_2$ the relation between the Ce-$O_{mig}$ bonds and the integrated COHP follows a straight line. The results for the 4fval model (Figure 9a) show modulation to the linear relation between M-$O_{mig}$ bond-length and the integrated COHP while the 4fcore model (Figure 9b) produces straight linearity. The modulation is due to competition between the ionic distance and the number of empty f orbitals in determining the strength of the interactions, which is turned off in the 4fcore model by fixing the occupancy of 4f orbitals. The analyses by scanning Ln elements as the dopant reveal that the variable occupancy of empty 4f orbitals is important typically for early Ln elements to produce the covalent interactions that essentially affect formation and migration of oxygen vacancies. Thus, for the group of higher atomic numbers with fewer numbers of empty 4f orbitals, i.e. Pm with $4f^4$ to Gd with $4f^7$, the 4f core model can be a convenient and efficient approximation in describing formation and migration of oxygen vacancies doped $CeO_2$ systems, related to the oxygen ion conduction.

**Conclusions**

In summary, systematic analysis by DFT+$U$ calculations was conducted to rigorously study the specific roles of 4f-states in Ln-doped $CeO_2$ systems. It was found that the 4f-electrons contribute to ionic/covalent interactions and implied a dual character of the 4f-electrons: strong localization on the ion, and delocalization by hybridization with oxygen 2p orbitals that contributes to covalent interactions. The covalent interactions become prominent at early Ln elements due to the number of empty 4f orbitals in the lone ion. The covalent interactions are facilitated by large hybridization due to $O_{2p}$ states which have large spatial extent reaching into the atomic sphere around the Ce atom so that $Ce_{4f}$ orbitals get delocalized by admixture with the character of $O_{2p}$. As an example, in Pr-doped $CeO_2$, it was demonstrated that the variable occupancy of empty 4f orbital of dopant determines the preferential site of oxygen vacancy formation agreeing with the EXAFS analysis reported in the experiment.[19] Thus, this work proves that contribution of 4f electrons to covalent interactions can be essential in systems comprising early lanthanide elements. Typically, it is emphasized that the treatment of 4f electrons with variable occupancy is crucial both in the host and dopant sites for producing correct properties of formation and migration of oxygen vacancies.

**Acknowledgements**

This Ph.D study was financially supported by Directorate General of Higher Education (DGHE) of Ministry of Research and Technology and Higher Education of Republic Indonesia. Research grant through Marubun Research Promotion Foundation (FY2015) is greatly acknowledged. This work supported in part by MEXT Grant-in-Aid for Scientific Research (15H05736, 24246013,

15KT0062, 26248006); "JST ACCEL Program "Creation of the Functional Materials on the Basis of the Inter-Element-Fusion Strategy and their Innovative Applications"; NEDO Project "R&D Towards Realizing an Innovative Energy Saving Hydrogen Society based on Quantum Dynamics Applications"; and the Osaka University Joining and Welding Research Institute Cooperative., Some of the numerical calculations presented here done using the computer facilities at the following institutes: the super computer centers of Institute of Solid State Physics (ISSP) of the University of Tokyo and Yukawa Institute for Theoretical Physics (YITP) of Kyoto University, High Energy Accelerator Research Organization (KEK) under a support of its Large Scale Simulation Program (No. 12/13-10), Cybermedia center (CMC) of Osaka University and the National Institute for Fusion Science (NIFS).

**Electronic Supporting Information**

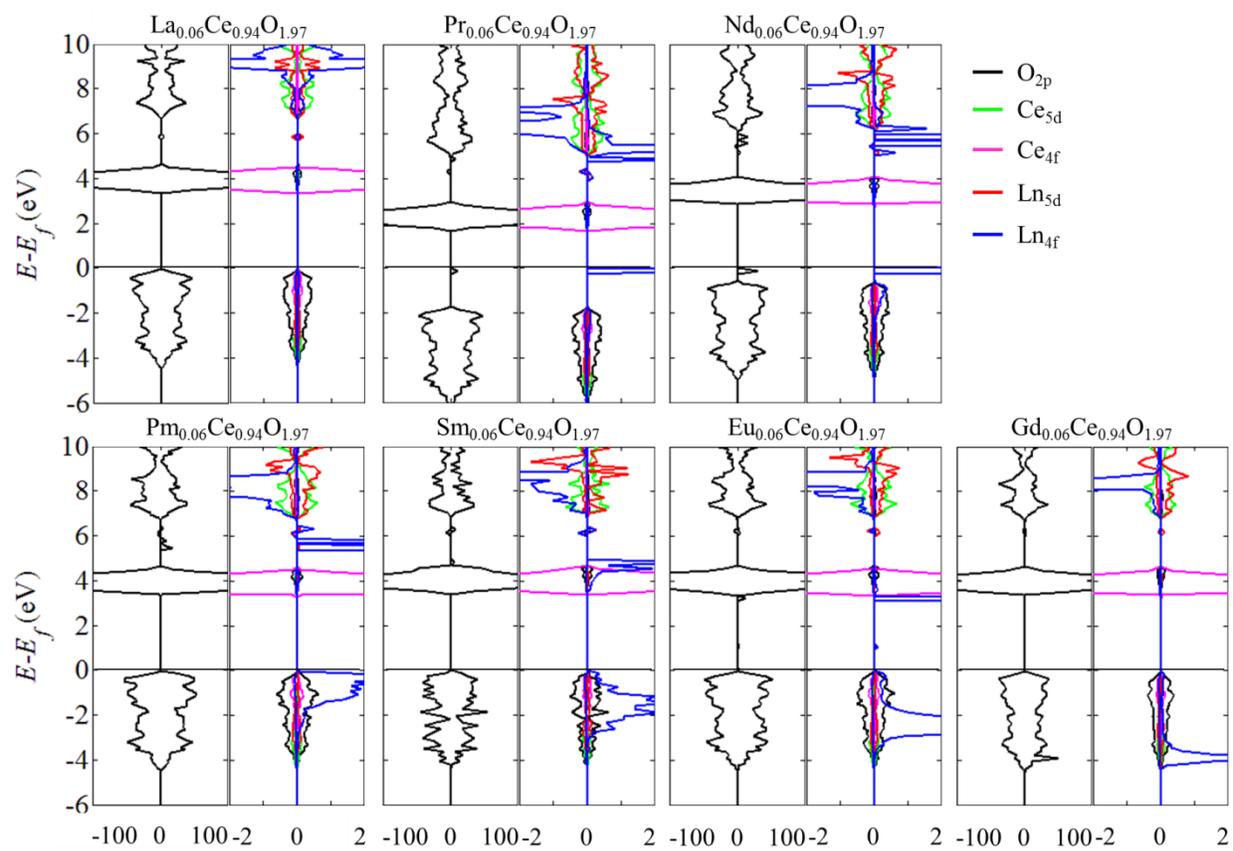

Figure S1. Total DOS and projected DOS obtained from hybrid functional calculations.

Table S1. The vacancy formation energies, $E_f^{vac}$, and migration energies, $\Delta E_m$, for ip1, ip2, and ip3. The calculations were conducted by considering 4f electrons frozen at the core ($Ln_{4f@core}$) and as valence electrons with $U_{eff@Ln} = 0$ eV as well as with the fitted $U_{eff@Ln}$ specified in the computational details section in the main manuscript.

| Systems | $Ln_{4f}$ Treatment | $\Delta E_m$ (eV) | | | $E_f^{vac}$ (eV) |
| --- | --- | --- | --- | --- | --- |
| | | ip1(fwd/bwd) | ip2(fwd/bwd) | ip3(fwd/bwd) | |
| LCO | $U_{eff@La} = 0$ eV | 0.338/0.424 | 0.811/0.782 | 1.591/1.591 | -0.614 |
| | $U_{eff@La} = 4.5$ eV | 0.305/0.465 | 0.794/0.811 | 1.695/1.695 | -0.662 |
| $CeO_{2-x}$ | $U_{eff@Ce}* = 5$ eV | | 0.442/0.442 | | 3.693 |
| | $Ce_{4f@core}**$ | 0.370/0.486 | 0.798/0.809 | 1.629/1.629 | 4.546 |
| PCO | $U_{eff@Pr} = 0$ eV | 0.534/0.333 | 0.395/0.330 | 0.968/0.968 | 3.127 |
| | $U_{eff@Pr} = 4$ eV | 0.466/0.414 | 0.810/0.746 | 1.361/1.361 | 1.640 |
| | $Pr_{4f@core}$ | 0.437/0.464 | 0.801/0.771 | 1.356/1.356 | -0.590 |
| NCO | $U_{eff@Nd} = 0$ eV | 0.536/0.366 | 0.700/0.662 | 0.898/0.898 | 1.077 |
| | $U_{eff@Nd} = 3.5$ eV | 0.497/0.431 | 0.788/0.747 | 1.325/1.325 | -0.100 |
| | $Nd_{4f@core}$ | 0.501/0.447 | 0.799/0.740 | 1.373/1.373 | -0.608 |
| PmCO | $U_{eff@Pm} = 0$ eV | 0.585/0.373 | 0.831/0.697 | 1.171/1.171 | 0.683 |
| | $U_{eff@Pm} = 3.5$ eV | 0.544/0.391 | 0.827/0.718 | 1.293/1.293 | -0.581 |
| | $Pm_{4f@core}$ | 0.574/0.430 | 0.793/0.700 | 1.256/1.256 | -0.656 |
| SCO | $U_{eff@Sm} = 0$ eV | 0.591/0.392 | 0.776/0.654 | 1.001/1.001 | 0.360 |
| | $U_{eff@Sm} = 4$ eV | 0.578/0.425 | 0.814/0.710 | 1.193/1.193 | -0.622 |
| | $Sm_{4f@core}$ | 0.612/0.404 | 0.788/0.669 | 1.189/1.189 | -0.697 |
| ECO | $U_{eff@Eu} = 0$ eV | 0.571/0.389 | 0.836/0.714 | 1.068/1.068 | -0.293 |
| | $U_{eff@Eu} = 4$ eV | 0.598/0.435 | 0.839/0.730 | 1.230/1.230 | -0.650 |
| GCO | $U_{eff@Gd} = 0$ eV | 0.723/0.337 | 0.775/0.575 | 1.015/1.015 | -0.636 |
| | $U_{eff@Gd} = 4$ eV | 0.708/0.355 | 0.774/0.590 | 1.039/1.039 | -0.793 |
| | $Gd_{4f@core}$ | 0.734/0.363 | 0.766/0.576 | 1.018/1.018 | -0.793 |
| YCO | -- | 0.818/0.316 | 0.750/0.503 | 0.906/0.906 | -0.891 |

\* $U_{eff} = 5$ eV for all Ce 4f-electrons.

\*\* The 4f-electrons of two reduced Ce ions as virtual dopants in the presence of an oxygen vacancy, are set as core electrons.

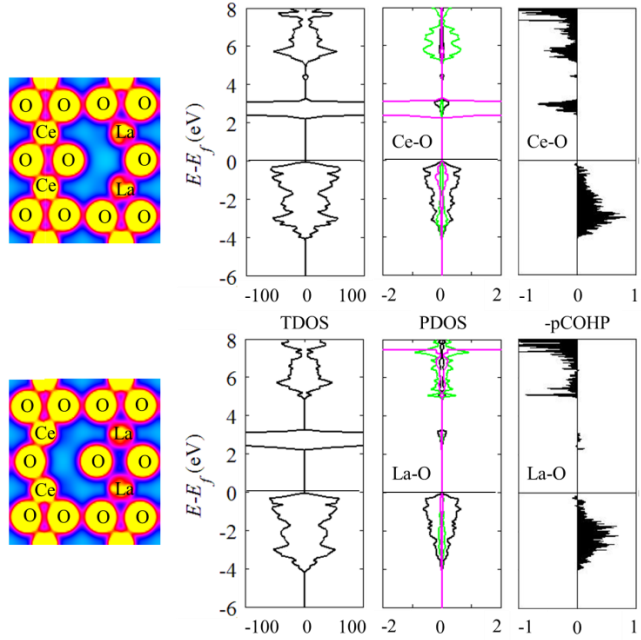

Figure S2. Charge density plots within energy range of $O_{2p}$-VB, total and projected DOS and the corresponding COHP curves of La-doped $CeO_2$ at initial and final states of ip1.

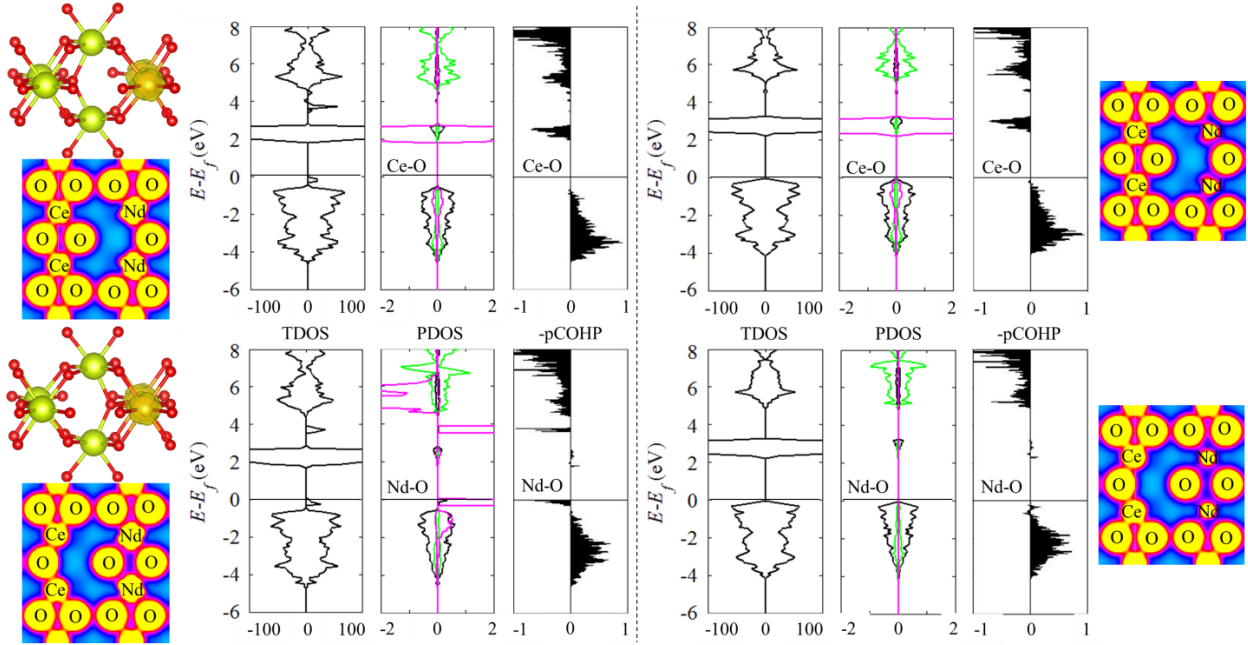

Figure S3. Charge density plots within energy range of $O_{2p}$-VB, total and projected DOS and the corresponding COHP curves of Nd-doped $CeO_2$ at initial and final states of ip1. Left panel shows the 4fval results while right panel shows the 4fcore results.

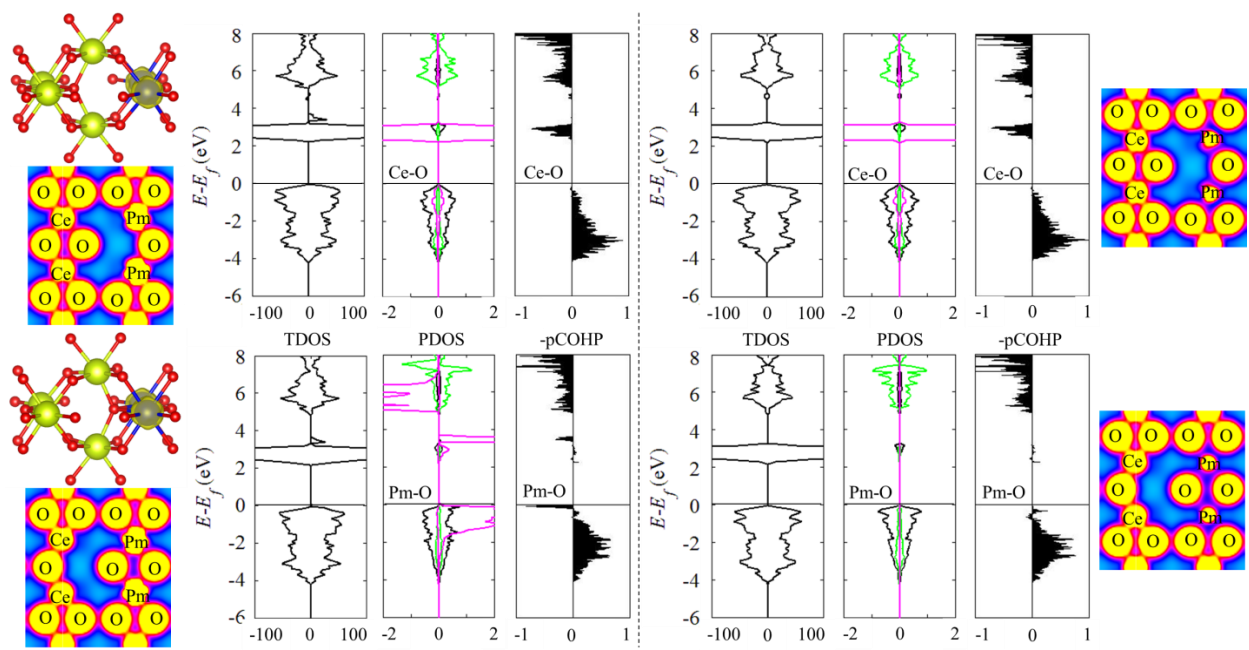

Figure S4. Charge density plots within energy range of $O_{2p}$-VB, total and projected DOS and the corresponding COHP curves of Pm-doped $CeO_2$ at initial and final states of ip1. Left panel shows the 4fval results while right panel shows the 4fcore results.

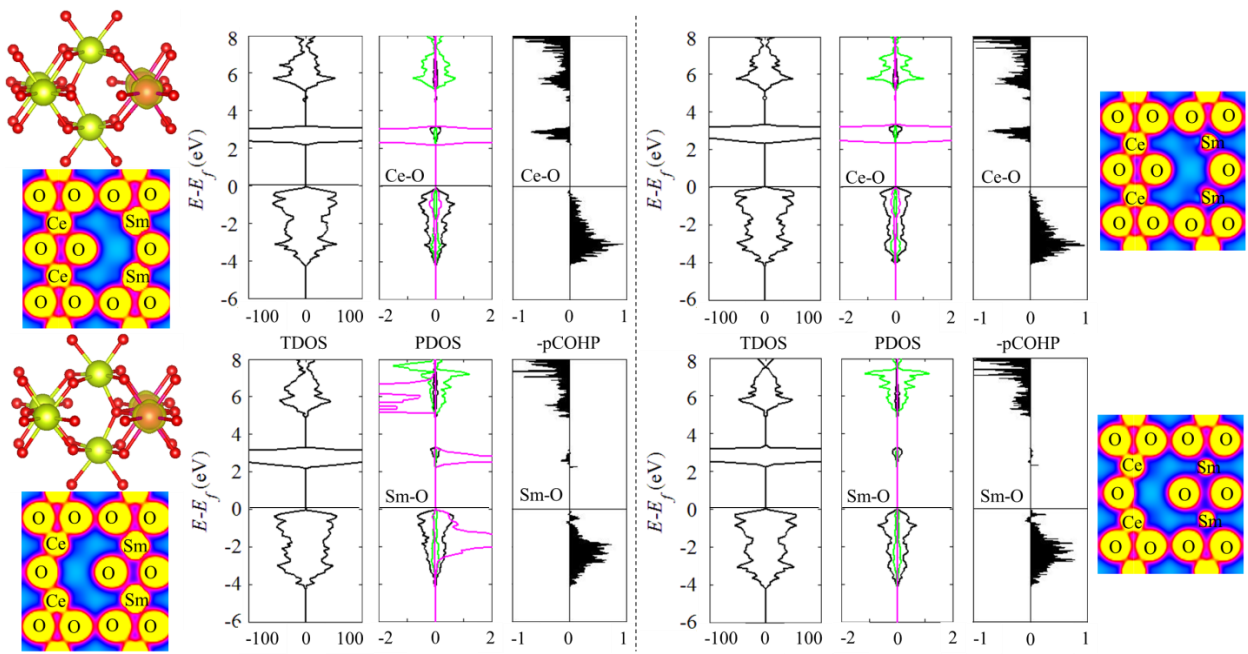

Figure S5. Charge density plots within energy range of $O_{2p}$-VB, total and projected DOS and the corresponding COHP curves of Sm-doped $CeO_2$ at initial and final states of ip1. Left panel shows the 4fval results while right panel shows the 4fcore results.

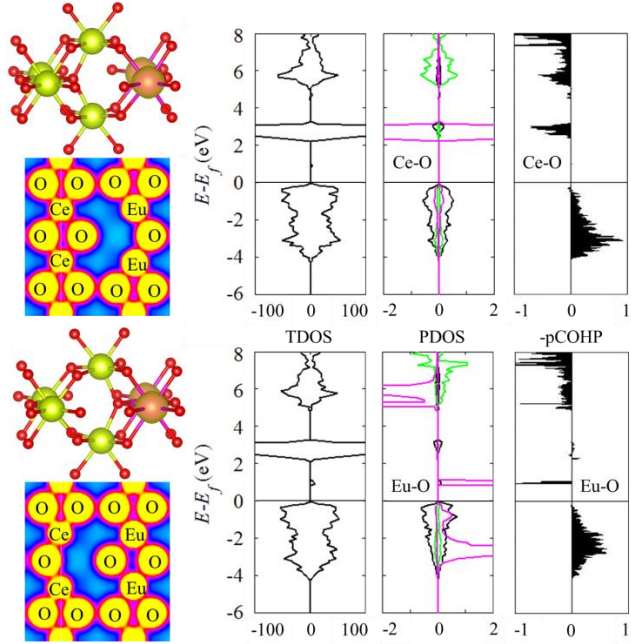

Figure S6. Charge density plots within energy range of $O_{2p}$-VB, total and projected DOS and the corresponding COHP curves of Eu-doped $CeO_2$ at initial and final states of ip1.

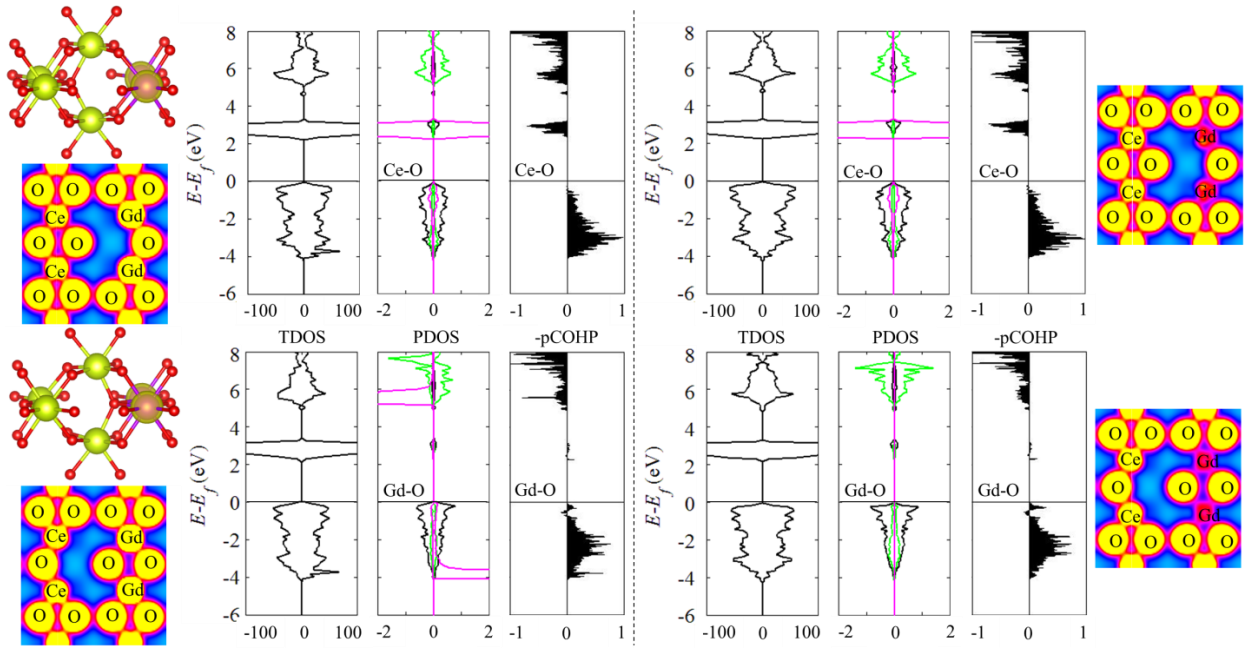

Figure S7. Charge density plots within energy range of $O_{2p}$-VB, total and projected DOS and the corresponding COHP curves of Gd-doped $CeO_2$ at initial and final states of ip1. Left panel shows the 4fval results while right panel shows the 4fcore results.

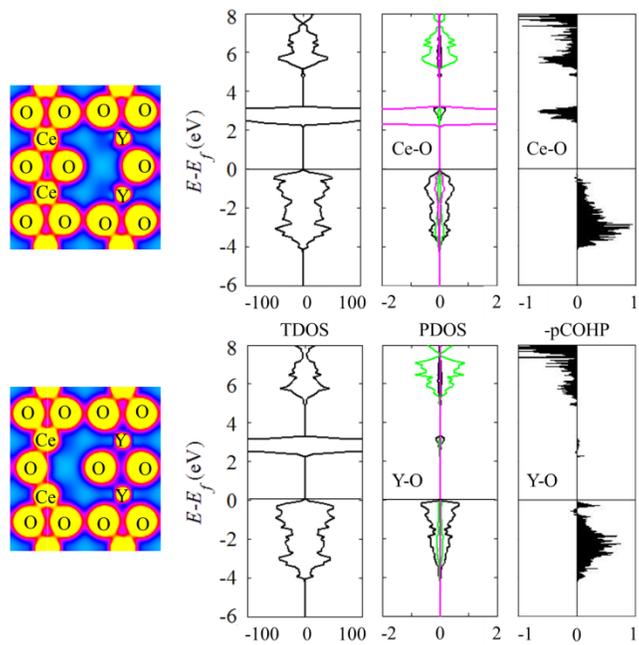

Figure S8. Charge density plots within energy range of $O_{2p}$-VB, total and projected DOS and the corresponding COHP curves of Y-doped $CeO_2$ at initial and final states of ip1.